\documentclass{aa}
\usepackage[varg]{txfonts}
\usepackage{graphicx}       
\usepackage{amsfonts,amsmath}
\usepackage{natbib}
\usepackage{color}
\newcommand{\be}{\begin{equation}}
\newcommand{\ee}{\end{equation}}
\newcommand{\f}{\frac}
\newcommand{\nn}{\nonumber}
\newcommand{\ord}{\mathcal{O}}
\newcommand\T{\rule{0pt}{2.6ex}}       
\newcommand\B{\rule[-1.2ex]{0pt}{0pt}} 
\newcommand{\M}{\mathcal{M}_c}

\begin{document}
\title{
A crystal ball for kilonovae
}
\subtitle{Early warnings from Einstein Telescope for electromagnetic follow-ups}
\author{Sarp Akcay\inst{1}\inst{2}
\and Morgan Fraser\inst{3}
\and Antonio Martin-Carrillo\inst{3}}

\institute{Theoretisch-Physikalisches Institut, Friedrich-Schiller-Universit{\"a}t, 07743, Jena, Germany
\and School of Mathematics \& Statistics, University College Dublin, Belfield, Dublin 4, Ireland
\and School of Physics, University College Dublin, Belfield, Dublin 4, Ireland} 
\abstract{
The discovery of gravitational waves from merging compact objects has opened up a new window to the Universe. Planned third-generation gravitational-wave detectors such as Einstein Telescope will potentially deliver hundreds of such events at redshifts below $z\sim0.1$. Finding electromagnetic counterparts to these events will be a major observational challenge. We demonstrate how Einstein Telescope will provide advance warning of 
such events on a timescale of hours, based on the low frequency emission from the pre-merger system. In addition, we suggest how this early warning enables prompt identification of any electromagnetic counterpart using the Large Synoptic Survey Telescope.
}
\maketitle

\section{Introduction}

With the first direct detection of gravitational waves (GWs) in 2015 by the Advanced Laser Interferometer Gravitational-Wave
Observatory (Advanced LIGO; \citealp{FirstGW}), gravitational-wave astronomy moved from prospect to reality. The first GW source observed by Advanced LIGO, GW150914, matched the signal predicted for the merger of two black holes with masses 36 and 29 M$_{\odot}$.

Only two years after GW15014, 
the Advanced LIGO and Virgo gravitational-wave observatories detected GW170817, with a waveform consistent with the merger of two neutron stars \citep{GW170817}. A spatially and temporally coincident short Gamma Ray Burst (GRB) was also seen by the {\it Fermi} and {\it INTEGRAL} satellites \citep{GW170817_GRB}. This discovery sparked a global effort to find the counterpart of GW170817 at optical wavelengths, which resulted in the identification of AT2017gfo less than 11 hours later \citep{GW170817_EM}. AT2017gfo faded exceptionally rapidly, and displayed cool temperatures and lines from unusual r-process elements at exceptionally high velocities \citep{Smar17,Arca17,Pian17,Coul17,Kilp17}. These characteristics marked AT2017gfo as a kilonova; a transient powered by the radioactive decay of short-lived nuclides formed in the merger of two neutron stars.

The detection of electromagnetic (EM) counterparts to GWs from merging neutron stars is of exceptional significance for astrophysics. Kilonovae are the predominant site for r-process nucleosynthesis, and so play a critical role in the chemical evolution of galaxies. If a kilonova can be identified and associated with a host galaxy of known redshift, then the degeneracy between inclination angle and distance inherent to a GW signal can be broken. This in turn allows the opening angle of the GRB jet to be constrained, something that has been done for a handful of GRBs to date \citep{2018ApJ...857..128J}. GW 
sources can also be used to independently determine the Hubble constant $H_0$ \citep{GWH0}.

The identification of AT2017gfo as the counterpart to GW170817 was realised by the 
ability of Advanced LIGO-Virgo to localise the GW signal to $\sim30~\mathrm{deg}^2$.
In addition, at only 40 Mpc, GW170817 was exceptionally close. This enabled the EM counterpart to be identified through targeted observations of galaxies at this distance within the GW localisation region \citep{Coul17}. Unfortunately, such a strategy is only feasible for the nearest GW sources, and rapidly becomes irrealisable beyond $\sim100$ - $200$ Mpc, both as the number of galaxies within the search volume increases, and as the fraction of galaxies with reliable redshifts decreases. 
This embarrassment of riches is a serious obstacle for identifying EM counterparts to GW transients that will be detected by Einstein Telescope in the 2030s \citep{ET_doc}.

 Einstein Telescope (ET) will consist of three
 V-shaped interferometers which eliminate blind spots and further allow it to construct a null
 stream \citep{Sathyaprakash:2012jk} which can be used to veto spurious events \citep{Wen:2005ui}. 
 Additionally, ET will be a xylophone \citep{Hild:2009ns}, i.e., a multi-band detector capable of delivering high sensitivities both at low frequencies ($\sim 5\,$Hz) and high frequencies ($\sim 100\,$Hz). 
 Here, we focus on the C configuration (ET-C) which offers the highest low-frequency sensitivity as shown in Fig.~\ref{fig:ETB2030}.
 ET-C will detect $\gtrsim\mathcal{O}(10^3)$ binary neutron star inspirals per year out to 1~Gpc \citep{Akcay18}.  
 A subset of these sources will be close enough that they will be detected a few hours
before their respective mergers \citep{Akcay18}, hence opening up the possibility of alerting EM
observatories to conduct follow-up observations \emph{before, during} and after the prompt gamma-ray burst. 
Additionally, ET-C will yearly forecast a few tidal disruption events from NS - BH (black hole) binaries \citep{Akcay18}.


To fully exploit the prospect of multi-messenger astronomy, a number of wide-field survey telescopes are either operational, in commissioning, or under construction. Foremost among these is the Large Synoptic Survey Telescope \citep[LSST;][]{LSSTbook}, which has an 8.4\,m primary mirror, and will image $9.6~\mathrm{deg}^2$ in a single pointing. Construction of LSST is well underway
with full survey operations starting in 2023.
Apart from LSST, the majority of current and next-generation survey telescopes have a relatively small mirror, but a large camera, and are designed to observe $\sim10-50~\mathrm{deg}^2$ in a single pointing to a limiting magnitude of $\sim20$ - $22$. ZTF \citep{ZTF}, GOTO \citep{GOTO} and ATLAS \citep{ATLAS} are all currently operational, while BlackGEM is under construction \citep{BlackGEM}.

There has been a considerable amount of discussion in the literature as to the optimal strategy to identify an EM counterpart to future GW transients \citep{2016ApJ...820..136G, 2011MNRAS.415L..26C, 2016A&A...592A..82G, 2017ApJ...834...84C, 2014MNRAS.437..649S, 2016MNRAS.462.1085A, 2018arXiv181204051M}. In most cases however, a large number of candidates will be found within the search region, for which further spectroscopic follow-up observations will be required. This spectroscopic classification bottleneck will remain a problem into the foreseeable future.


Our aim here is to demonstrate exciting EM follow-up studies that can be done by taking advantage of the early GW warning capability of ET. 
The ability of ET to detect inspiraling compact objects $\sim$hours before the merger has been described in detail by \cite{Akcay18}, and in this Letter we 
present a brief summary of these results and their implications for follow-up observations of electromagnetic counterparts.

We use $f$ to denote the quadrupole GW frequency
in the detector frame. $G$ is Newton's constant and $c$ is the speed of light.

\section{Einstein Telescope}
\label{sect:et}
In this section, we compute the advance warning times ($T_\text{AW}$) ET will provide (computational details are provided in \cite{Akcay18}).
Consider a binary neutron star (BNS) system with component masses
$m_1, m_2$ inspiraling at a luminosity distance $D$ with a corresponding redshift $z$. For GW frequencies of interest here ($f \lesssim 10\,$Hz), the binary undergoes an adiabatic inspiral dominated by
the emission of leading-order (quadrupole)
gravitational radiation. By balancing the
power emission in GWs to the rate of change of binding energy, we obtain the frequency evolution of the GW frequency
\be
\dot{f} = \f{96}{5}\pi^{8/3} \f{(G \mathcal{M}_c)}{c^5}^{5/3}\, f^{11/3}, \label{eq:fdot}
\ee
where $\mathcal{M}_c  = (1+z){(m_1 m_2)^{3/5}}{(m_1+m_2)^{-1/5}}$ is the redshifted chirp mass. 
After fixing the integration constant, Eq.~(\ref{eq:fdot})
can be integrated to yield the time left to merger at a given frequency, usually called the inspiral time
\begin{align}
\tau_\text{insp}(f) &= \f{5}{256\pi}\f{c^5}{(\pi G \M)^{5/3}} \,f^{-8/3}\nn\\
&=16.72\,\text{minutes} \, \left(\f{1.219 M_\odot}{\M}\right)^{5/3}\,\left(\f{10\,\text{Hz}}{f}\right)^{8/3}
\label{eq:tau_insp}\, .
\end{align}
This result can be supplemented with a post-Newtonian series up to $\ord(c^{-2})$ \citep{Blanchet_LRR}, but the resulting expressions
 only change $\tau_\text{insp}$ by $\lesssim 2\%$.

A passing GW induces a response in a given interferometer (IFO) known as the GW strain. 
In frequency domain, the norm of the GW strain is given by
$|\tilde{h}(f)|=A h_0 f^{-7/6} |Q|,$ 
%
where $A= \pi^{-2/3}(5/24)^{1/2},\, h_0 = c  (1+z)^{-1}\tilde{M}^{5/6}/D$ with $\tilde{M}= G \M c^{-3}$ and $Q$ is the IFO quality factor which is a function of source sky location and inclination angles $\{\theta,\phi, \iota\}$, and the relative detector-source polarization angle $\psi$.

The IFO response to a GW strain is quantified in terms of a signal-to-noise ratio (SNR).
As we can not a priori know the angles $\{\theta,\phi,\iota,\psi\}$, we use an angle-RMS-averaged SNR, 
appropriate for a detector with triangular topology like ET
\be
\rho_{\text{ET}}(f_1,f_2) = \frac{3}{5}\,2 A\, h_0  (1+z)^{-1/6} \left[\int_{f_1}^{f_2} d f'\, \f{f'^{-7/3}}{S_{n}(f')}\right]^{1/2} \label{eq:ET_SNR},
\ee
where 
$\sqrt{S_n(f)}$ is the {\it amplitude spectral density} of the detector (also called detector noise) and
the factor of $3/5$ in Eq.~(\ref{eq:ET_SNR})
is due to RMS-averaging \citep{Akcay18}. 
For a given source, $\rho_\text{ET}$ may vary by $\sim \pm 30\%$ compared to the average (\ref{eq:ET_SNR}), but will always be $>0$ because of ET's all-sky coverage.

We define $T_\text{AW}$ as the time left to merger from the moment of detection 
defined as follows: let $f_0$ be the frequency
at which the GW strain equals the RMS-averaged detector noise, i.e., $\tfrac{5}{3}\sqrt{S_n(f_0)}=2\sqrt{f_0} \tilde{H}_\text{ET}(f_0)$ where $\tilde{H}_\text{ET}(f)= h_0 f^{-7/6}$;
the instant of detection is given by 
$\bar{f}>f_0$ such that $\rho_\text{ET}(f_0,\bar{f})=15$. 
Thus $T_\text{AW} = \tau_\text{insp}(\bar{f})$.
The total accumulated SNR is given by $\rho_\text{tot}=\rho_\text{ET}(f_0, f_\text{ISCO})$, 
where $f_\text{ISCO}$ is the frequency at which the inspiral transitions to plunge. 
Here, we use the standard approximation from general relativity: 
$f_\text{ISCO} \approx {c^3}\left[{6^{3/2}\pi\, G (m_1+m_2)(1+z)}\right]^{-1} \simeq {1571}(1-z) \left(\frac{2.8M_\odot}{m_1+m_2}\right)\text{Hz}$ for $z\ll 1$. 
We chose a conservative threshold SNR of 15. 

\begin{figure*}
\sidecaption
\includegraphics[width=12cm]{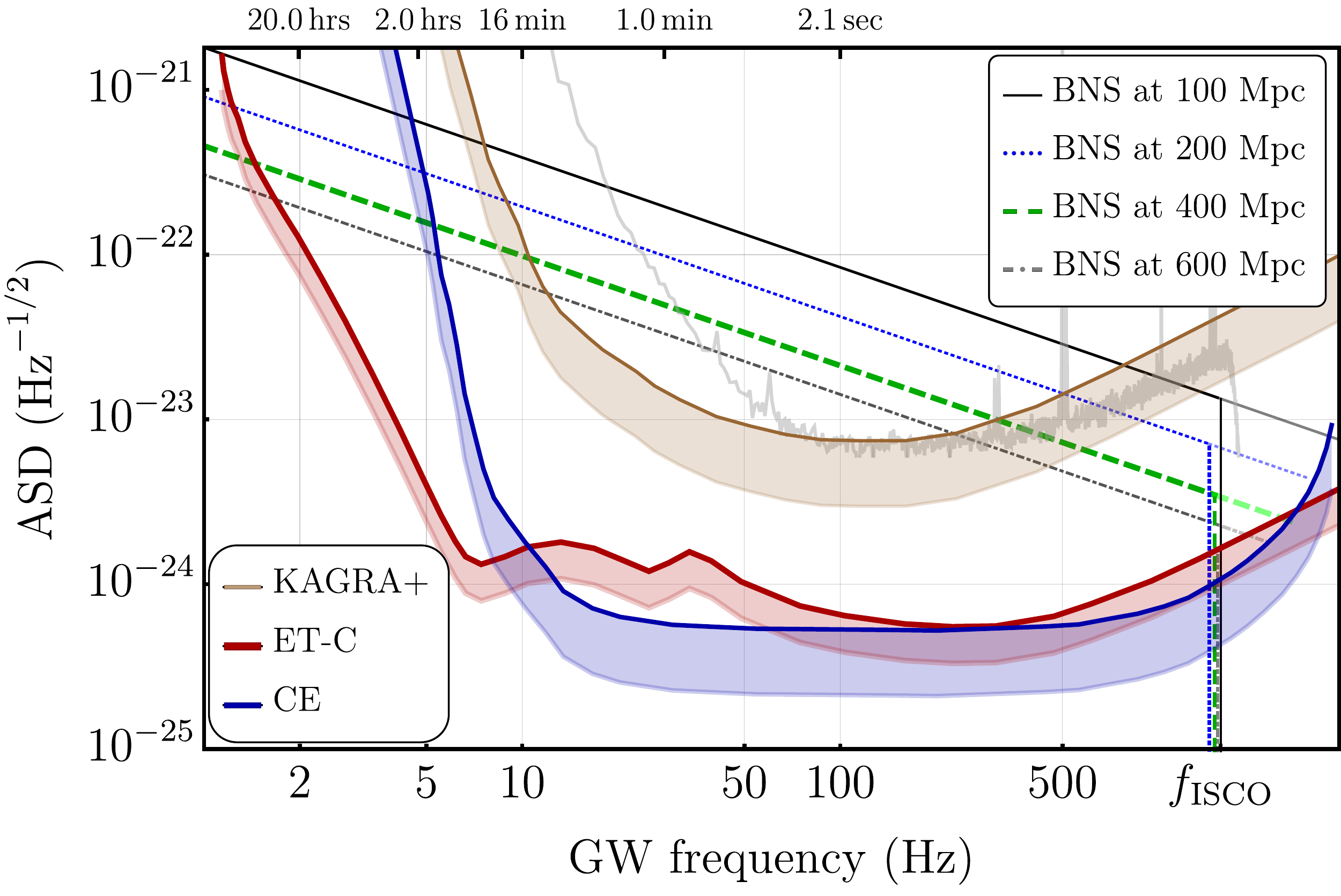}
\caption{
$1.4 M_\sun$ - $1.4 M_\sun$ inspiralling BNS systems sweeping across the sensitivity band of Einstein Telescope's C configuration (thick red curve).
The solid (black), dotted (blue), dashed (green), and dot-dashed lines (gray) lines are the redshift-corrected GW strains, $2\sqrt{f}\tilde{H}_\text{ET}$, at luminosity distances of $D=100, 200, 400, 600\,$Mpc, respectively. 
The vertical lines with correspondingly identical patterns (colors) mark the redshifted ISCO frequencies $(1+z)^{-1} f_\text{ISCO}$ at which point we terminate each inspiral.
As the true ISCO frequency is likely larger than $f_\text{ISCO}$ \citep{Marronetti:2003hx}, the inspirals would continue to nearly 2\,kHz indicated by the faded lines in the plot.
We also show the sensitivity curves for Cosmic Explorer (blue) and KAGRA+ (brown) with the solid curves representing their RMS-averaged sensitivities and the bottom of each shaded region, the maximum sensitivities.
The faint gray curve represents the sensitivity of Advanced LIGO during GW170817.
The upper  horizontal axis gives the time to merger for a BNS at 100 Mpc.
}
\label{fig:ETB2030}
\end{figure*}

$D$ now fully determines $T_\text{AW}$.
In Fig.~\ref{fig:ETB2030} we display the GW strain for four canonical ($m_1=m_2=1.4 M_\odot$) BNS inspirals at $D=100,200, 400, 600\,$Mpc, 
compared to the expected ET-C sensitivity.
Note that the inspiral GW strains scale as $ \sqrt{f} f^{-7/6} = f^{-2/3}$ \citep{Colpi_Sesana},
hence are straight lines with slope $=-2/3$. 
The frequencies $f_0$, at which each inspiral enters ET-C's sensitivity band, are the intersection
points between these straight lines and the ET-C noise and lie within 1 - 2\,Hz.
We list the advance warning times along with the total SNRs for these sources in  Table~\ref{table:ET}, where
we show that ET-C is capable of providing up to \emph{five} hours of early warning before a  merger.
%

%
%
%
%
\begin{table}[h]
\caption{Forecasting capabilities of the C configuration of Einstein Telescope. 
$D$ is the luminosity distance and $z$ the corresponding redshift 
computed assuming a flat $\Lambda$CDM universe with $\Omega_\Lambda = 0.6911, \Omega_m = 0.3089, H_0 = 67.74~\mathrm{km}~\mathrm{s}^{-1}~\mathrm{Mpc}^{-1}$ \citep{Planck2015}.
$\bar{f}$ is the threshold frequency at which ET-C accumulates SNR of 15.
$T_\text{AW}\equiv \tau_\text{insp}(\bar{f})$ [see Eq.~(\ref{eq:tau_insp})].
$\rho_\text{tot}$ is the total SNR for each inspiral.
Last column lists the event rates 
within concentric shells of radius $D$ and thickness 100\,Mpc.}
\label{table:ET}
\centering
\begin{tabular}{lccccc}
\hline\hline
$D\,$(Mpc) &  $z$ & $\bar{f}\,$(Hz) &  $T_\text{AW}\,$(hrs)& ${\rho}_\text{tot}$ &$ R\,(\text{yr}^{-1})$ \T\B \\
\hline
100 & $ 0.022$ & 3.3 & 5.3 & 365 & $1.54^{\,+3.20}_{\,-1.22}$ \T \B \\
200 & $ 0.044$ & 4.1 & 2.9 & 182 & $10.8^{\,+22.4}_{\,-8.54}$ \T \B \\
300 & $ 0.065$ & 4.66 & 1.9 & 121 & $29.3^{\,+60.8}_{\,-23.2}$ \T \B \\
400 & $ 0.085$ & 5.1 & 1.5 & 90.5 & $57.0^{\,+118}_{\,-45.1}$ \T \B \\
500 & $ 0.10$ & 5.4 & 1.2 & 72.2 & $93.9^{\,+195}_{\,-74.4}$ \T \B \\
600 & $ 0.12$ & 5.7 & 1.03 & 60.0 & $140^{\,+291}_{\,-111}$ \T \B \\
\hline\hline
\end{tabular}
\end{table}
%
%
%
%
%
%
%
\subsection{Event rates for binary neutron star inspirals}
We base our event rate calculations on $R=1540^{+3200}_{-1220}\,\text{Gpc}^{-3}\,\text{yr}^{-1}$ inferred from GW170817 \citep{GW170817} consistent with $110 - 3840~\mathrm{Gpc}^{-3}\,\text{yr}^{-1}$ \citep{LIGOScientific:2018mvr}. 
This translates roughly to $R(D\le 600\,\text{Mpc})=330^{+690}_{-260}~\mathrm{yr}^{-1}$. 
We partition this volume into concentric shells each 100 Mpc thick. For each shell, we show the event rates in Table~\ref{table:ET}. 
%
%
%
%

We see from Table~\ref{table:ET} that 
within $z\lesssim 0.1$, 
ET-C will annually detect GWs from $\sim$ 40 to 600 BNS transients.
Roughly 7\% of these will be within 200\,Mpc yielding $T_\text{AW}\gtrsim 3\,$hours which means that EM observatories will be alerted $\sim 3$ - 40 times yearly with the prospect of observing births of kilonovae.
If we assume that future EM observatories could respond with $T_\text{AW}=1\,$hour warning, then
all BNS transients within 600\,Mpc become possible candidates for merger-kilonovae observations.
In this case, early warnings would be sent by ET at least once a week and at most three times per day.

\subsection{Source localisation estimations}
ET will have poor source localisation because all three of its IFOs will be co-located. 
However, as ET-C will
detect BNS inspirals hours before the merger, during this time the Earth will have rotated by $\sim 10^\circ$ - $50^\circ$.
Therefore, within 1Gpc, ET \emph{alone} will localise  $\sim20$\% of sources to within $100~\mathrm{deg}^2$, 
2\% to within $20~\mathrm{deg}^2$, and 0.5\% to within $10~\mathrm{deg}^2$ \citep{Zhao:2017cbb} 
\footnote{\cite{Zhao:2017cbb} perform their calculations for ET-D which in fact has slightly worse sensitivity for $f\lesssim 5\,$Hz than ET-C, cf. Fig.~19 of \cite{GW_IFO_LRR}}.
Rescaling to $D=400$~Mpc, corresponding to $T_\text{AW}= 1.5\,$hrs, we obtain $\gtrsim 5$ yearly BNS transients with $\Delta\Omega \lesssim 10\,$deg$^2$ and $\gtrsim20$ transients with $\Delta\Omega \lesssim 20\,$deg$^2$.


Our above approximations on localisation should be taken as the pessimistic case as 
ET will be not be operating alone. 
Currently proposed companion detectors are the Japanese cryogenic detector KAGRA \citep{Akutsu:2017thy,KAGRA2}, LIGO's successor Voyager \citep{LIGO_Voy}, its ``relative'' IndIGO \citep{Unnikrishnan:2013qwa}, 
and finally, Cosmic Explorer (CE): the third generation US detector with 40\,km arm length \citep{Evans:2016mbw}.  
Assuming a three-detector network consisting of ET and two CEs with a total network SNR $> 12$ and two detectors 
each with SNR $> 5$, \cite{Mills:2017urp} find that half of signals will be localised to within $1~\mathrm{deg}^2$
out to a redshift of $z\sim 0.25$. However, this survey is not concerned with issuing sufficiently early warnings to EM facilities. 
The problem is that only ET has the extreme low-frequency sensitivity enabling $T_\text{AW}\sim\,$hours.
The other detectors will not accumulate any SNR from BNS inspirals in the $f\lesssim 5\,$Hz domain\footnote{There is a prospect for improving LIGO's low-frequency end called LIGO-LF \citep{Yu:2017zgi}.}.
However, CE will be sensitive enough to accumulate SNR from 5\,Hz for BNSs with $D\lesssim 400\,$Mpc (Fig.~\ref{fig:ETB2030}).
Given that its sensitivity increases steeply between 5 and 10\,Hz, CE will accumulate $\text{SNR} = 5$ with 1.5\,hours left to merger and SNR = 15 with 1.25\,hours left
resulting in a total network SNR, $\rho_\text{net} \equiv (\rho_\text{ET}^2+\rho_\text{CE}^2)^{1/2}=\{18.8,27.4\}$ 
for $\tau_\text{insp}=\{1.5, 1.25\}\,$hours, respectively.
As localisation improves with increasing SNR, this means that an initial $\Delta\Omega$ of $\sim 100\,$deg$^2$
can be reduced as the BNS inspirals through its second to last hour before the merger.

This region can be further decreased once the BNS enters a third detector's bandwidth,
which will most likely be the mid-2020s-upgraded KAGRA+ with its strain sensitivity shown as the brown curve in Fig.~\ref{fig:ETB2030}.
Using the same analysis as for CE, we find that KAGRA+ will pick up a 400-Mpc inspiral at $\sim 10\,$Hz and will accumulate $\text{SNR}=5$ within a minute. 
For a 100-Mpc source, KAGRA+ would accumulate $\text{SNR} > 15$ more than a half hour before the merger.
Thus, for nearby sources, even KAGRA+ sensitivity could contribute to pre-merger localisation efforts. Given that LSST requires $\sim 5\,$minutes to
point anywhere in the sky, KAGRA's contribution will matter.
Once again, this is a rather pessimistic estimation as we expect the 2030s KAGRA
to be more sensitive than the brown curve of Fig.~\ref{fig:ETB2030}.

In short, within 400\,Mpc, we can annually expect five BNS transients to be localised to $10~\mathrm{deg}^2$ 1.5 hours
before the merger. We can expect an additional $\sim15$ more with initial localisation of $\sim 20~\mathrm{deg}^2$ 
which we expect will be narrowed down to $\sim10~\mathrm{deg}^2$ about one hour before the merger.
We envision a three-stage localisation procedure whereby ET conducts the operations alone until $f\sim 5\,$Hz 
--- roughly two hours before merger --- at which point CE joins in and, finally, around $f\sim 10\,$Hz KAGRA+ starts accumulating SNR with $\gtrsim 15\,$ minutes left to merger.

\section{Implications for optical follow-up of GW detections} \label{sect:EM}
Identifying an optical or near-infrared (NIR) counterpart to a GW is an observational challenge. If a GW is only localised to tens, or even hundreds of square degrees, then we must survey a large area of the sky to find an EM counterpart. While large format CCDs make taking imaging of an area of $\sim100~\mathrm{deg}^2$ relatively straightforward, we must identify our EM counterpart of interest among the many unrelated astrophysical transients that we expect by chance within the same area. Thus far, this has relied upon large scale efforts to spectroscopically classify credible candidates that are found within the sky localisation of a GW. For example, for the BH merger GW151226, \cite{Smar16} found 49 candidate transients within 290 deg$^2$, and obtained spectra for 20 of these. Such a survey strategy is clearly an inefficient use of scarce telescope time.

The early warning obtained for future GW events discussed in Sect. \ref{sect:et} offers an alternative approach for finding EM counterparts. In brief, if we can detect a GW with $\sim$1 hr advance warning, and can localise it to $\sim50~\mathrm{deg}^2$ or better, then we can obtain imaging of this area both immediately prior to, and after, the merger happens. Since the merger will be the only thing that has changed over such a short period of time, identifying an EM counterpart in difference imaging becomes straightforward.

While this section focuses on the optical/NIR part of the EM spectrum, 
we also considered the implications of an early GW warning at higher energies. Gamma-ray instruments have large field-of-views of the order of $1-2$~sr and a consequently high probability of detecting the prompt emission from the merger by chance. 
The detection of X-ray prompt emission (below 5 keV) could have significant scientific value, however most of the instruments sensitive at those energies have fields of view of the order of arcmin$^2$. Currently, the JEM-X instrument onboard INTEGRAL is the only instrument with a sufficiently large field-of-view ($25~\mathrm{deg}^2$). Of the missions currently under study, only THESEUS \citep{2018AdSpR..62..191A} would have an X-ray instrument with a large field-of-view (1 sr at $0.3-5$ keV). However, its sensitivity ($10^{-10}~\mathrm{erg}~\mathrm{cm}^{-2}~\mathrm{s}^{-1}$ in 1000 s) will be too low to detect the prompt emission of most NS-NS or NS-BH mergers.

\subsection{The rates and nature of contaminants}

There are broadly three classes of contaminants that we must consider when searching for EM counterparts to GWs: stellar variables and flares such as cataclysmic variables (CV); variability in Active Galactic Nucleii (AGN); and supernovae (SNe). The first class of contaminants show a strong dependence on Galactic latitude \citep{Drak14}, and are concentrated in the disk of the Milky Way. In addition, for at least some CVs, a quiescent counterpart will be visible in deep images, or in other cases prior outbursts may have been detected. We hence expect that CVs and other variable stars will be a relatively minor source of contamination for EM counterpart searches. This is further borne out by \cite{Smar16}, who found only 3 of 49 potential counterparts to GW151226 to be stellar.
AGN can often be identified through their historical lightcurves, which may show previous variability, or through the presence of a cataloged X-ray or radio counterpart. Given the relatively straightforward removal of stellar and AGN contaminants, we are left with SNe as the dominant contaminant. Again, in the case of GW151226, 88\% of potential counterparts turned out to be SNe. Within a magnitude limited survey around three quarters of supernovae detected will be Type Ia SNe \citep{2011MNRAS.412.1441L}, due to their luminosity. So, to first order, our main source of contamination when searching for EM counterparts of GW events will be SNe Ia.

\begin{figure}[h]
\includegraphics[width=\linewidth]{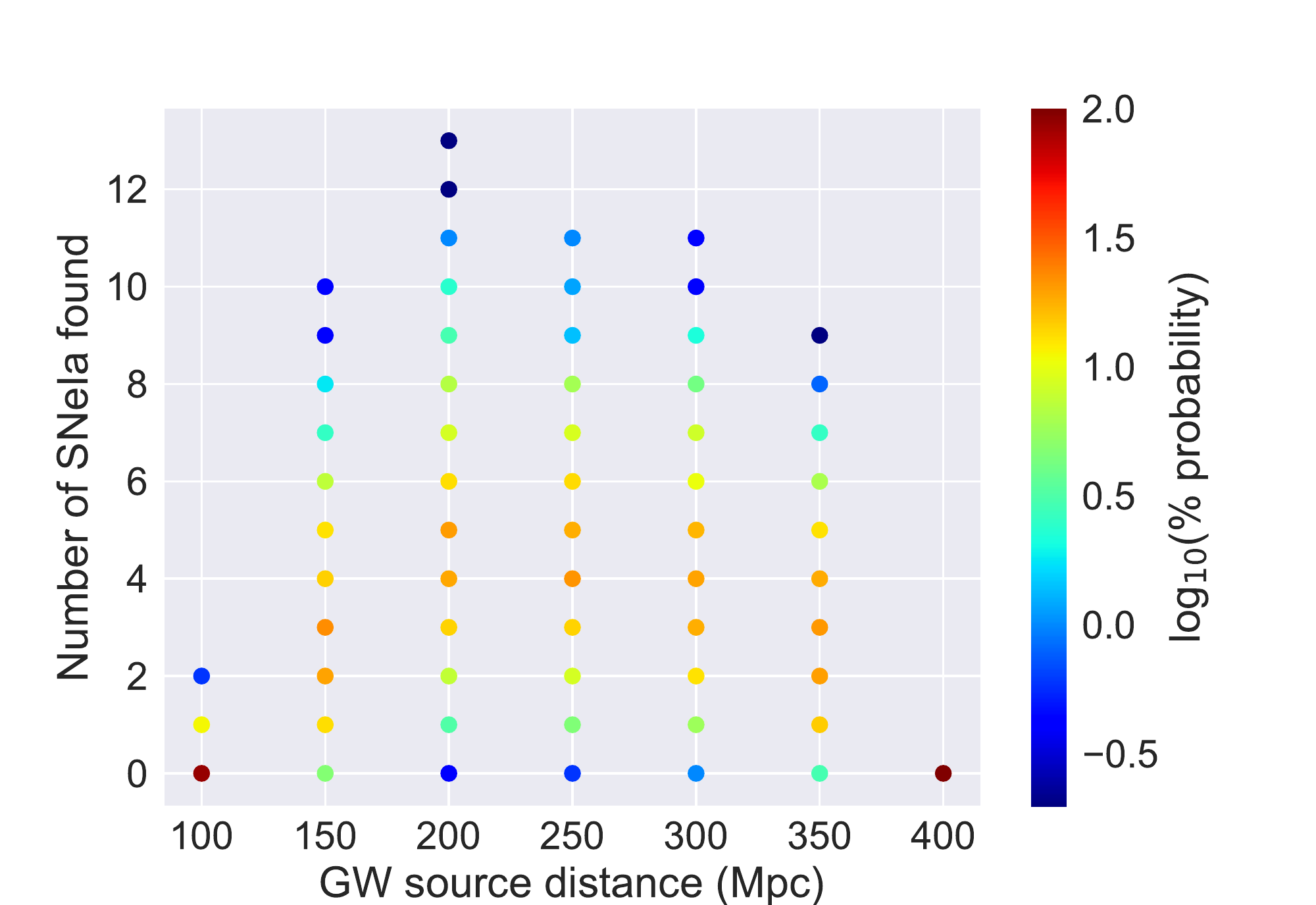}
\caption{The probability of finding a given number of SNe Ia (that are likely to be confused with a kilonova), as a function of distance. We assume a survey with a four night cadence reaching $m_r\lesssim$22, a GW source localised to 30 deg$^2$, and a kilonova  $M_r=-15$.}
\label{fig:SNIa}
\end{figure}

We ran Monte Carlo simulations to estimate of the number of unrelated SNe Ia that may be found in a search for an EM counterpart to a GW transient. We took the volumetric SN-Ia rate from \cite{2010ApJ...713.1026D}, and assumed that the GW event could be localised to a region of $\sim$30 deg$^2$. We further assumed an optical survey with a cadence of 4 days and a limiting magnitude $m_r\sim$22, and that the kilonova had an absolute magnitude of $M_r=-15$ at peak. We then calculated the number of SNe Ia that would be detected by the survey with a magnitude comparable ($\pm1$~mag) to the kilonova, {\it and} where the SN would have not been detected on the previous image taken of the field four nights earlier. The results of this are shown in Fig. \ref{fig:SNIa}, where we find that for a GW source at a distance of a few hundred Mpc, we will typically have four 
unrelated SNe Ia that are impossible to distinguish from a kilonova solely on the basis of single filter imaging. In the worst case scenarios, we may have as many as ten contaminants that are observationally similar to the kilonova.

This calculation should be taken as an approximate guide, and the exact numbers of contaminants will vary with a number of factors. Higher-cadence transient surveys, and better localisation of GW events will reduce the number of contaminants found. On the other hand, simply increasing the depth of a survey will not necessarily improve matters; while some SNe will be detected earlier (and hence ruled out) when their lightcurves are still rising, greater numbers of more distant SNe will also be detected. In any case though, it is reasonable to expect that even in the most favourable scenarios we will always have more than a single plausible candidate counterpart to any future GWs.

\subsection{A proposed strategy for EM counterpart detection}

We have demonstrated that unrelated SNe Ia will be found within the region to which a GW is localised, and that a few of these are going to appear as new sources with comparable magnitude to a potential kilonova. There are two main reasons why this is a problem. Firstly, obtaining a spectrum of a kilonova and $\sim$5 unrelated SNe Ia will take commensurately longer.
Secondly, unrelated transients make it much harder to obtain spectra of potentially rapidly fading BH-NS mergers, or of their very early ($<$1 hr) evolution. If we have five sources of comparable magnitude, and it takes $\sim1$ hr to obtain a spectrum of each, we will only obtain a spectrum of the correct candidate in $<$1~hr in a minority of cases.
The early warning from ET offers a solution to both of these problems. We propose that as soon as sufficient SNR is accumulated to localise a GW to $\sim20~\mathrm{deg}^2$ or better, we take a set of images for that region of the sky {\it before} the merger occurs.

The LSST can reach a $5\sigma$ limiting magnitude of $r\sim24.3$ with $2\times15$~s images. The readout times for each exposure will be 2~s, and as the field-of-view of LSST is $9.6~\mathrm{deg}^2$, we will likely be able to cover the entire GW footprint in a small number of pointings. Even allowing 5 minutes for slewing of the telescope, it is likely that obtaining pre-merger images will take at most 10 minutes.
After the merger, we would take a second set of images. These would be subtracted from the pre-merger template images taken $\lesssim1$ hr before; and the kilonova should be the only thing that has changed over this brief period, making it trivial to identify.

\section{Summary}

We have shown that Einstein Telescope will provide $\sim$~hr early warning of BNS mergers,
and in a substantial fraction of cases will localise these to better than $100~\mathrm{deg}^2$.
This warning provides sufficient time to trigger observations of the region with wide-field
optical telescopes, {\it immediately prior} to the merger. These images can then be used as templates
to compare a second set of post-merger images, allowing for the rapid identification of any possible
EM counterpart, {\it without} the need for costly spectroscopic screening. In addition, this technique allows for 
counterparts to be found faster, which may be essential to follow up the most rapidly fading sources.


\begin{acknowledgements}
 S.~A. acknowledges support by the EU H2020 under ERC Starting Grant, no.~BinGraSp-714626.
 MF is supported by a Royal Society - Science Foundation Ireland University Research Fellowship.
\end{acknowledgements}

\bibliographystyle{aa}
\bibliography{GWbib}

\end{document}